\documentclass[%
 reprint,
nofootinbib,
amsmath,
amssymb,
aps,
]{revtex4-2}

\usepackage{graphicx}
\usepackage{dcolumn}
\usepackage{bm}
\usepackage[hyperfootnotes=false]{hyperref}
\usepackage{multirow}
\usepackage[makeroom]{cancel}
\usepackage[export]{adjustbox}

\newcommand{\fig}[1]{Fig.~\ref{#1}}
\newcommand{\eq}[1]{Eq.~\ref{#1}}
\renewcommand{\sec}[1]{Sec.~\ref{#1}}
\newcommand{\ket}[1]{\mbox{$|#1\rangle$}}
\newcommand{\bra}[1]{\mbox{$\langle #1|$}}
\newcommand{\braket}[2]{\mbox{$\langle #1|#2\rangle$}}

\newcommand{\beq}{\begin{equation}}
\newcommand{\eeq}{\end{equation}}

\renewcommand{\ss}{Schr{\"o}dinger's equation}
\newcommand{\ntwo}{\frac{1}{\sqrt{2}}}

\begin{document}
\preprint{APS/123-QED}
\title{Beyond Copenhagen: Following the Trail of Decoherence\\ in Feynman's Light Microscope} 

\author{Brian C. Odom}
 \email{b-odom@northwestern.edu}
\affiliation{%
 Center for Fundamental Physics, Department of Physics and Astronomy, Northwestern University, Evanston, Illinois 60208, USA 
}%


\date{\today}

\begin{abstract}
Feynman's light microscope invites us to reconsider what we thought we knew about quantum reality. Rather than invoking wavefunction collapse to predict the loss of fringes in a monitored interferometer, Feynman analyzes the problem in terms of a disturbance. This approach raises the question of whether the classical world, including its localized particles and definite measurement outcomes, might emerge as the universe evolves smoothly according to Schr\"odinger's equation. Treating the particle and its environment as an entangled system, unmodified quantum mechanics shows remarkable success toward this end. This is the purview of decoherence theory. How we then think about macroscopic reality becomes dependent on how we think about microscopic reality. Is quantum mechanics successful because it describes what microscopic particles are really doing, such as traveling both interferometer paths at the same time? Or is the wavefunction only a mathematical tool which predicts measurement outcomes but does not describe microscopic reality? Both options are uncomfortable. The first implies that each moment in time branches into a vast number of divergent macroscopic realities. The second represents, for many practitioners, a weakened view of science. This article is written to be accessible to anyone with an undergraduate course in quantum mechanics.
\end{abstract}

\maketitle

\section{Introduction}
The Copenhagen interpretation is dying. At least, that is true for its variant most familiar to physicists. Microscopic reality was said to evolve smoothly according to \ss, but at an ill-defined macroscopic boundary the wavefunction collapsed. Physicists had long expressed suspicion of this ``shifty split"~\cite{bell1990against}. Nonetheless, until nearly the end of the 20th century, the distinction between small and big things seemed clear enough in practice, and the Copenhagen theory made remarkably successful predictions. 

Experimental and theoretical developments have gradually changed the situation. Improving technologies allow investigation of quantum behavior of increasingly large systems, calling into question any distinction between microscopic and macroscopic. Nanoclusters with nearly the mass and diameter of small viruses have created interference patterns~\cite{pedalino2025probing}, mechanical oscillators on the 100 picogram scale have been prepared in entangled states~\cite{kotler2021direct}, and mesoscopic fullerene molecules have been shown to decohere smoothly and predictably in the presence of background gas~\cite{hornberger2003collisional} and as they radiate thermal energy~\cite{hackermuller2004decoherence}. Meanwhile, decoherence theory~\cite{zeh1970interpretation, joos1985emergence, zurek1991decoherence} has demonstrated that environmental entanglement explains how many, if not all, features of classical physics emerge naturally from unbiased quantum mechanics.

\begin{figure}[htbp!]
    \centering
    \includegraphics[width=0.4\textwidth]{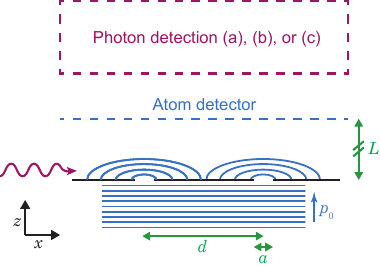}
    \caption{Light microscope apparatus. The origin for $x$ and $z$ is midway between the slits. }
    \label{slits}
\end{figure}

Feynman's light microscope thought experiment~\cite{feynmanlectures, feynmanonline} can be viewed as a prelude to these developments. An atom scatters a photon while traversing a two-slit interferometer (\fig{slits},) and the photon goes undetected. The Copenhagen theory says that we should not see an interference pattern if the photon has sufficiently short wavelength to have made possible a which-path measurement, because the atom wavefunction has collapsed into one of the classical single-path states. Feynman had the good sense not to treat such measurement postulates as inscrutable. Since the interaction is a quantum process, might we be able to understand the disappearance of fringes without invoking collapse? The light microscope provides a wonderful testing grounds for examining, both theoretically~\cite{wootters1979complementarity, scully1991quantum, tan1993loss, storey1994path, facchi2004thermal, drezet2006heisenberg, drezet2006momentum,stanford,lawrence2022observing} and experimentally~\cite{chapman1995photon, durr1998origin, kokorowski2001single, walborn2002double, cronin2009optics}, what lies beneath quantum measurement and how the classical world emerges from quantum physics. It is remarkable that such a fruitful idea originated from Feynman's lectures~\cite{feynmanlectures, feynmanonline} meant for first-year undergraduate students!

Feynman's semiclassical explanation~\footnote{Feynman warns that his rough introduction will need revision. Later, in a less well-known portion of the text, he explains fringe washout using entangled states.} is that fringes average out because the atom is perturbed by recoil. This analysis succeeds at showing the uncertainty principle prevents a logical crisis; it is impossible to observe an atom traveling a single path while also observing interference fringes indicating two-path traversal. However, semiclassical reasoning fails to explain why multiple photons scattered by the single-atom-wave will all originate from only one path or the other. It also fails to explain at a conceptual level what makes classical states special. Why does environmental disturbance seem to steer the atom toward behaving as if it takes a single path, when two-path traversal is equally legal in quantum mechanics?

By accounting for atom-photon entanglement, decoherence theory addresses both semiclassical shortcomings. Acknowledging that we are usually ignorant of some aspect of the entangled environment, we can predict the emergence of tame classical states from the wildly free quantum world. The wavefunction decoheres into non-interfering classical storylines, each with a different set of facts. As a result, the Copenhagen shifty split is eliminated. Using the Born rule to assign probabilities, we can make predictions without ever imposing any distinction between between microscopic and macroscopic.

Those predictions have so far passed all experimental tests. However, post-Copenhagen quantum mechanics leaves no comfortable option for interpreting reality. Do the equations of quantum mechanics describe some reality outside our minds? If so, then we are currently experiencing only one of many parallel branches of reality, with most other branches looking very different from ours. We could instead insist that, out of the predicted wavefunction branches, the only real one is that corresponding to our experience. But then we would need to concede that the wavefunction does not describe microscopic reality, that it is just a mathematical tool for predicting ``detector clicks." Alternatively, we might be dissatisfied with either option and suspect that quantum mechanics needs some fundamental modification, which could change the discussion entirely.

\section{The Setup}
The apparatus (\fig{slits}) has a wall in the $z=0$ plane and slits of width $a$, centered at $\pm \frac{d}{2} \hat{x}$. There is an atom detector at $L\hat{z}$. The atom wavepacket approaches the slits from below with central momentum $p_0 \hat{z}$, and passage through the slits does not change this longitudinal momentum. The diffracted atom does acquire from the wall a distribution of transverse momenta $p_x$. Just as it leaves the slits, we can represent the atom in the single-path basis 
\beq
\label{psii}
\ket{\psi_\text{i}}=\ntwo\left(\ket{\text{L}}+\ket{\text{R}}\right),
\eeq
where $\ket{\text{L}}$ and $\ket{\text{R}}$ describe wavepackets localized at either slit, and $\braket{\text{L}}{\text{R}}=0$.

\begin{figure}[htbp!]
    \centering
    \includegraphics[width=0.47\textwidth]{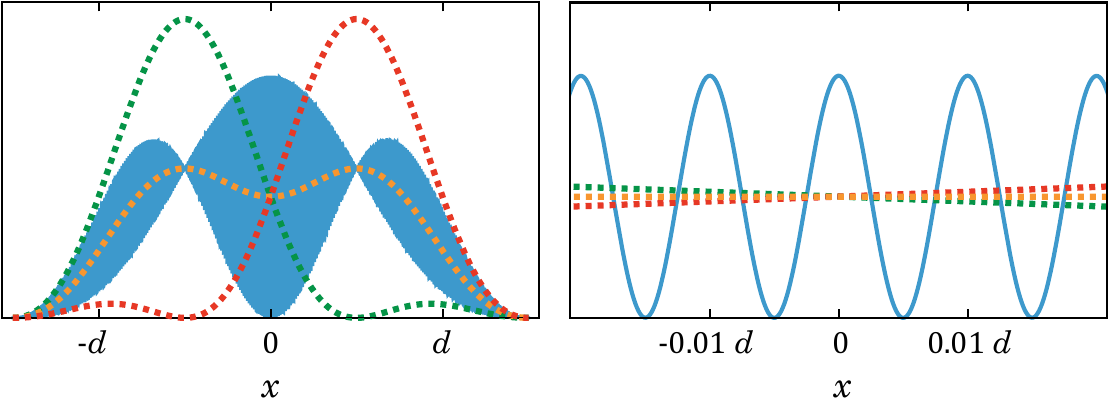}
    \caption{Normalized atom probability at $z=L$, without a scattered photon. Green/red: single-slit traversal. Orange: incoherent sum of single-slit sources. Blue: coherent two-slit traversal. The right panel shows a zoomed-in view such that fringes are resolved. Fringe minima are not generally at zero because the atom detector is not in the far-field, so the two slits contribute different amplitudes except at $x=0$.}
    \label{nophoton}
\end{figure}

The most common configuration is to place the atom detector in the far-field, where calculation of interference patterns is simpler. But then the single-slit diffraction patterns are indistinguishable from each other and also from a decohered two-slit pattern. For our purposes, it is worth the calculational trouble to bring the atom detector closer so that those three patterns are distinct, and we use $L=10\,d$. Atom patterns are shown in \fig{nophoton}, and the procedure for calculating them described in Appendix~\ref{appCalc}.

The region just above the slits is illuminated with light of wavelength $\lambda$ and $\vec{k}=k \hat{x}.$ For simplicity, we consider only $\lambda \gg a$, so that an atom passing through one slit is approximately a point source for scattered photons. The atom recoil from the elastic scattering is $\hbar \vec{\kappa}=\hbar \vec{k}-\hbar\vec{k}_\text{f}$, where $\hbar \vec{k}_\text{f}$ is the momentum of the scattered photon. Note that the light is not focused, so both sites contribute an amplitude for each scattered photon. (Whether or not those amplitudes are balanced for any detected photon will turn out to depend on the wavelength and detection configuration.)

We wish to explore the short-wavelength case $\lambda = d/10$, while also keeping $\lambda \gg a$. This leads our choice $a=d/100$, which has the inconvenient feature that there a large number of fringes ($\sim100$) within each single-slit envelope (\fig{nophoton}.)

\begin{figure}[htbp!]
    \centering
    \includegraphics[width=0.45\textwidth]{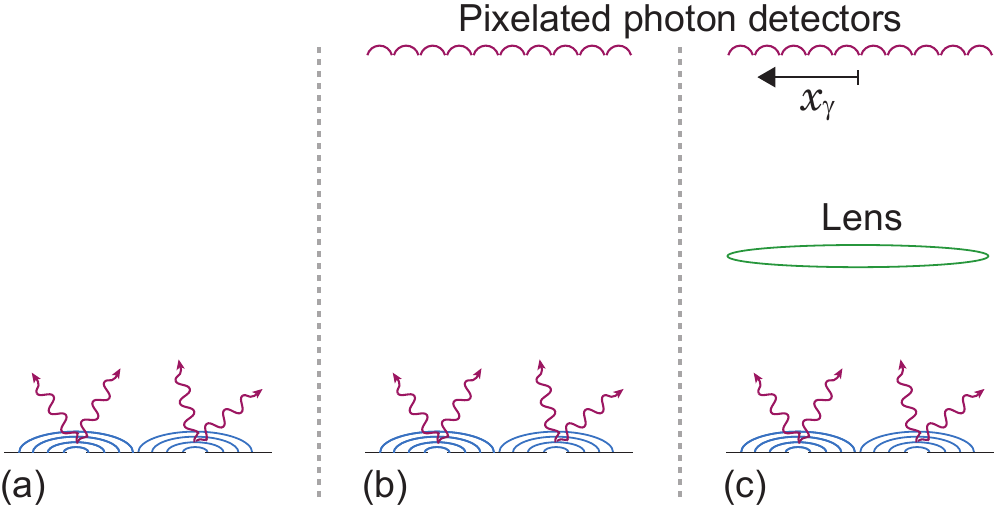}
    \caption{Different photon detection configurations (not to scale) to complete Fig.~\ref{slits}. Red wavy arrows represent light scattered from the atom. (a) No photon detection. (b) Far-field photon detection. (c) Imaging detection. (This simplest lens configuration causes images to be inverted, so we define $x_\gamma=-x$.)}
    \label{photonDetection}
\end{figure}

We either ignore the scattered photons, or detect them in one of two configurations (\fig{photonDetection}.) The first has a photon detector in the far-field so that each pixel corresponds to a well-defined $\vec{k}_\text{f}$. The second configuration uses a lens to image each position $x$ at $z=0$ onto position $x_\gamma=x$ on the detector plane; this is a microscope configuration.

\section{Anticipating Decoherence}
\label{anticipating}
Feynman offers a semiclassical explanation for why fringe washout always accompanies which-path information becoming available. ``By trying to `watch' the [atoms] we have changed their motions. That is, the jolt given to the [atom] when the photon is scattered by it is such as to change the [atom’s] motion enough so that if it might have gone to where [the probability] was at a maximum it will instead land where [the probability] was a minimum; that is why we no longer see the wavy interference effects~\cite{feynmanlectures, feynmanonline}." 

However, it is not enough to know that atom deflection is of order a fringe; we must also know the interferometer phase after recoil. Feynman implies that for the two-slit interferometer, an atom deflected to a new angle somehow carries along its original interferometer phase. We show in App.~\ref{appSC} that this is correct, so Feynman's semiclassical argument indeed predicts fringe washout for $\lambda \lesssim d$. (But for a counterexample, see a three-grating-interferometer experiment where fringe contrast remains high even when atoms are deflected by many fringes~\cite{chapman1995photon,cronin2009optics}.)

Although it is less intuitive, there is a more straightforward way to obtain the same result. To describe the recoiling atom wavefunction, we increase each momentum component of $\psi_\text{i}(\vec{r})$ by $\hbar\vec{\kappa}$, such that $\psi(\vec{r}) = e^{i \vec{\kappa} \vec{r}} \psi_\text{i}(\vec{r})$. Before recoil, the atom wavefunction leaving the left and right slit shared a common phase. After recoil, these phases differ by $\kappa_x d$, so the atom fringe locations depend on photon scattering angle. Since $0< \kappa_x < 2\times2\pi / \lambda$, averaging over angles washes out fringes if $\lambda \lesssim d$. 

Both arguments successfully predict fringe washout when short-wavelength photons are scattered. But by neglecting entanglement, they are limited in two ways. They suggest that decoherence can be blamed on perturbations to the atom as it traverses both paths. If this were the whole story, then with $\lambda \ll d$ and higher light intensity, we should be able observe an atom scattering a first photon from the left path and then a second photon from the right path. The atom wavefunction would not appear to collapse into a position eigenstate after a position measurement, thus contradicting the Copenhagen prediction and experiments~\cite{cronin2009optics}. More generally, these semiclassical arguments do not begin with the underlying quantum mechanics and predict emergence of classical behavior.

\section{Including Entanglement}
In our limit $\lambda \gg a$, the momentum imparted by the photon is much less than the spread imparted by the slit confinement. We can then make the approximation that the state of an atom passing through a single slit is not changed by photon scattering. For instance, for left-slit traversal
\beq
\ket{\chi} = \ket{\text{L}}\ket{\gamma_\text{L}}, 
\label{oneslitchange}
\eeq
where $\ket{\chi}$ is the post-recoil combined atom-photon state, \ket{\text{L}} is the pre-recoil atom state, \ket{\gamma_\text{L}} is a photon scattered from the left slit, and we have absorbed any phase factor into $\ket{\gamma_\text{L}}$. 

Above, \ket{\chi} was a non-entangled product state. The story is very different when we also allow passage through the other slit. The light beam crosses both slits, so each atom has an amplitude to scatter from both sites:
\beq
\label{joint}
\ket{\chi} = \ntwo \Bigl(\ket{\text{L}}\ket{\gamma_\text{L}} + \ket{\text{R}}\ket{\gamma_\text{R}}\Bigr).
\eeq
Now $\ket{\chi}$ is significantly entangled if $\lambda \lesssim d$, because then the photon resolves the slits, and $\ket{\gamma_\text{L}}\not\approx\ket{\gamma_\text{R}}$. For single-slit traversal, we could speak separately of an atom state and a scattered photon state. But for two-slit traversal and $\lambda \lesssim d$, we have no choice but to speak of a joint atom-photon state.

We can now see how an entanglement treatment correctly predicts that a single atom is never observed to scatter short-wavelength photons from both paths. If we scatter two photons, then \eq{joint} becomes
\beq
\label{jointTwo}
\ket{\chi_2} = \ntwo \Bigl(\ket{\text{L}}\ket{\gamma_\text{L}}\ket{\gamma_\text{L}} + \ket{\text{R}}\ket{\gamma_\text{R}}\ket{\gamma_\text{R}}\Bigr).
\eeq
Since $\braket{\text{L},\gamma_\text{L},\gamma_\text{L}}{\chi_2}\neq 0$, we might see two photons coming from the left path. But  $\braket{\text{L},\gamma_\text{R},\gamma_\text{L}}{\chi_2}\propto \braket{\gamma_\text{R}}{\gamma_\text{L}}$, which vanishes for $\lambda \ll d$~\cite{tan1993loss}. So, we never see one photon obviously scattered from the left path and the other from the right. 

\section{Different Photon Bases}
\label{bases}
We also want to represent the scattered photons in different bases. From \eq{joint},
\beq
\ket{\chi}=\ntwo\sum_n\Bigl[\braket{\gamma_n}{\gamma_\text{L}}\ket{\text{L}}+\braket{\gamma_n}{\gamma_\text{R}}\ket{\text{R}}\Bigr] \ket{\gamma_n}.
\label{chiDecomp}
\eeq
The photon momentum basis allows us to analyze the problem in terms of definite-recoil atoms, and it is also the basis we need when considering photons striking the far-field detector of Fig.~\ref{photonDetection}(b). In this basis, we have
\beq
\ket{\chi}=\ntwo\sum_{\vec{k}_\text{f}} \Bigl[e^{-i \kappa_x d/2}\ket{\text{L}}+e^{i \kappa_x d/2}\ket{\text{R}}\Bigr]S(\vec{k}_\text{f})\ket{\vec{k}_\text{f}},
\label{chiP}
\eeq
where $S(\vec{k}_\text{f})$ describes the scattering amplitude from an atom located at the origin~\cite{tan1993loss, drezet2006heisenberg, drezet2006momentum}. The relative phase $\kappa_x d=(\vec{k} - \vec{k_{\text {f}}})\cdot d\hat{x}$ can be understood as expressing the same local photon phase at each slit in terms of the $\vec{k}_\text{f}$ mode, offset by the phase accumulated by the incoming photon in traveling between the slits. This interferometer phase shift $\kappa_x d$ for each definite-recoil state was anticipated in the semiclassical discussion.

The basis of \eq{joint} is always useful for treating photons as originating from definite locations. However, just because the photon has a crisp origination position does not mean it will arrive at a crisp position $x_\gamma$ on the imaging detector of Fig.~\ref{photonDetection}(c). For considering these images, we use 
\beq
\ket{\chi}=\ntwo\sum_{x_\gamma}\Bigl[\braket{x_\gamma}{\gamma_\text{L}}\ket{\text{L}}+\braket{x_\gamma}{\gamma_\text{R}}\ket{\text{R}}\Bigr] \ket{x_\gamma}.
\label{chiX}
\eeq
Analytic expressions for the coefficients are given in~\cite{tan1993loss}, but we already know from classical optics some of their qualitative features. For $\lambda \ll d$ and a good lens, images of scattering from the two slits will be well-resolved, and $\braket{x_\gamma}{\gamma_\text{L}}$ will be zero if $\braket{x_\gamma}{\gamma_\text{R}}$ is non-zero. For $\lambda \gg d$, the images will be completely unresolved and $\braket{x_\gamma}{\gamma_\text{L}} \approx \braket{x_\gamma}{\gamma_\text{R}}$.

\section{Decoherence Basics}
If the atom probability distribution matches that of two non-interfering sources (orange curve in \fig{nophoton},) then each individual atom is said to have decohered. Classical interferometers have various imperfections, such as spectrally broad sources, which can lead to fringe washout. But in the quantum context, the term `decoherence' is generally reserved for that arising from entanglement. As we shall see, there are important aspects of (quantum) decoherence which have no classical analogy.  \emph{The first ingredient for decoherence is environmental entanglement}. In the light microscope, the photons are the environment. \emph{The second ingredient for decoherence is ignorance.} Decoherence is what happens when we have lost track of some part of an entangled state. 

\begin{figure*}[htbp!]
    \centering
    \includegraphics[width=0.95\textwidth]{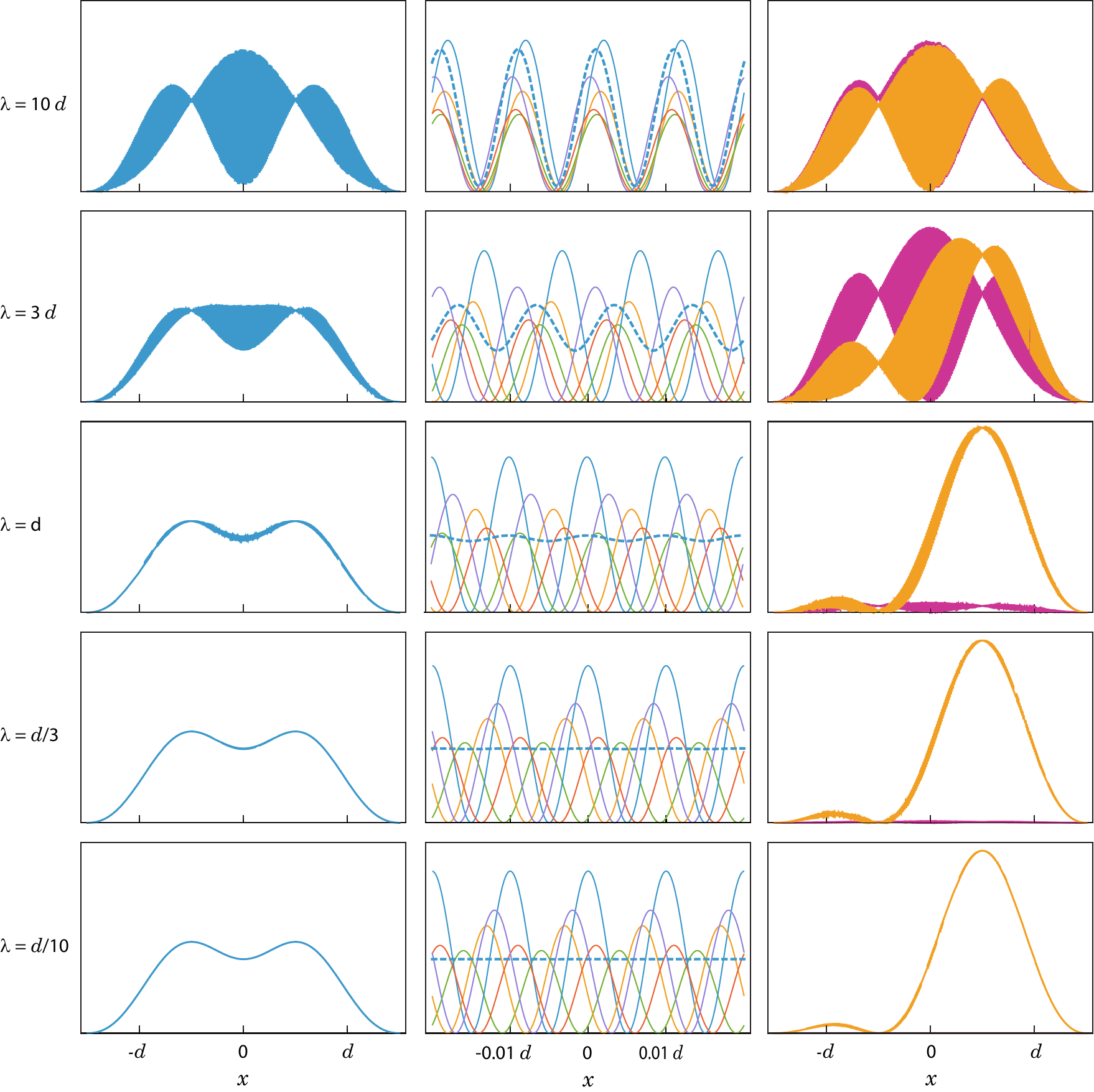}
    \caption{Atom probability patterns for different photon detection configurations, with fringes unresolved in the left and right columns. Left: No photon detection. Middle: Far-field photon detection. Solid lines show the atom probability patterns for selected photon scattering angles, corresponding to the photon detected on various pixels. (What is actually plotted is the arbitrarily scaled joint atom-photon probability as a function of $x$, for various $x_\gamma$.) Dashed lines show the average over all photon angles. Right: Imaging detection. Plots show the joint atom-photon probability as a function of atom detection at position $x$, for the cases of photon detection at $x_\gamma=d/2$ (orange) and at $x_\gamma=0$ (magenta).}
    \label{partials}
\end{figure*}

One way for decoherence to occur in the light microscope is for us to simply fail to set up a photon detector, as in \fig{photonDetection}(a). We then obtain the atom pattern by summing probabilities over possible photon states. Using Eq.~\ref{joint}, the probability of detecting the atom at some position $\vec{r}$ is given by 
\begin{align}
\label{P}
\mathcal{P}(\vec{r}) 
  &= \left|\braket{\vec{r}, \gamma_\text{R}}{\chi} \right|^2 + \left|\braket{\vec{r}, \gamma_\text{L}}{\chi} \right|^2\\
  &= \frac{1}{2} |\psi_\text{R}(\vec{r})|^2+ \frac{1}{2} |\psi_\text{L}(\vec{r})|^2+\text{Re}\left[\psi_\text{R}^*(\vec{r}) \psi_\text{L}(\vec{r}) \braket{\gamma_\text{R}}{\gamma_\text{L}}\right] \nonumber.
\end{align}
It is now clear that decoherence is intimately tied to entanglement. If $\lambda\ll d$, the scattered photon states are quite different from one another, and $\braket{\gamma_\text{R}}{\gamma_\text{L}}\approx0$~\cite{tan1993loss}. In this case, the atom and photon are maximally entangled in \eq{joint}, and \eq{P} predicts complete fringe washout. If $\lambda\gg d$, then $\braket{\gamma_\text{R}}{\gamma_\text{L}}\approx1$, \eq{joint} is nearly factorizable (non-entangled), and we predict survival of interference fringes even though a photon was scattered. The progression of decoherence, as the undetected photon wavelength decreases, can be seen in the left column of \fig{partials}.

The Born rule (used in \eq{P}) by itself predicts that scattering of an undetected short-wavelength photons leads to fringe washout.  These photons have potential to provide which-path information to a properly configured detector. There is overlap here with a commonly taught version of the Copenhagen theory, which says that the wavefunction collapses if a which-path measurement could potentially be performed, regardless of whether such a measurement is actually performed. \eq{P} shows that decoherence theory reveals the physics underlying the potentially/actually aspect of this Copenhagen postulate. In any case, there is no need for a macroscopic photon detector for fringe washout to occur.

\section{If We Do Detect Photons}
In fact, if we do have a photon detector, then the `ignorance' criterion for decoherence can be compromised. By recording atoms only corresponding to photons detected in some state, we can recover the ``partial interference patterns"~\cite{wootters1979complementarity, tan1993loss}. One set of these patterns drives home the point that underlying coherence is never truly lost, even under conditions where it would typically be hidden by our ignorance.

Using the far-field photon detection setup of \fig{photonDetection}(b), atom patterns correlated with selected $\vec{k}_\text{f}$ can be recorded. Each of these patterns corresponds to one term in the sum of \eq{chiP}. These terms represent phase-shifted atom interferometers, and atom patterns for various photon scattering angles are shown in the middle column of \fig{partials}. Scattering a $\lambda \ll d$ photon does not in itself lead to decoherence, as demonstrated by atom patterns for every $\vec{k}_\text{f}$ still having high-contrast fringes. In terms of measurement, fringe contrast preservation here is because photons landing on the far-field detector are in definite-momentum states; since these states are plane waves with infinite spatial extent, they do not resolve the slits and carry no which-path information. A three-grating version of this experiment showed the expected fringe recovery from otherwise washed-out atom patterns~\cite{chapman1995photon,cronin2009optics} and is a demonstration of a ``quantum eraser"~\cite{scully1991quantum}. Finally, we see in \fig{partials} that averaging atom patterns over all photon angles results in atom fringe washout only if $\lambda \lesssim d$, consistent with the results we obtained using the photon-origination basis in \eq{P}.

On the other hand, if we use the imaging setup of \fig{photonDetection}(c), we expect that detection of photons \emph{will} provide which-path information, provided $\lambda \lesssim d$. Atom patterns corresponding to photon detection at $x_\gamma=d/2$ are drawn in orange in the right column of \fig{partials}. For $\lambda\gg d$, the interferometer's relative amplitude and phase in \eq{chiX} are not modified significantly by the photon overlap integrals, so the atom fringes remain unchanged. In terms of measurement, this is because the slit images are unresolved, and each photon could equally likely have come from either slit; there has been no which-path information obtained by detecting these photons. In the regime $\lambda\gtrsim d$, $|\braket{x_\gamma=d/2}{\gamma_\text{R}}| > |\braket{x_\gamma=d/2}{\gamma_\text{L}}|$. We still have interference fringes, but the photon is beginning to distort the interferometer amplitudes, causing the atom pattern to shift rightward. For $\lambda \ll d$, $\braket{x_\gamma=d/2}{\gamma_\text{L}}\approx 0$. This limit represents a full-resolution which-path measurement, and the atom pattern corresponds to passage through only the right slit. (Note that the residual small bump at $x<0$ is due to the single-slit diffraction pattern, rather than any passage through the left slit.) \fig{partials} also shows in magenta atom patterns for photons detection at $x_\gamma=0$. Atoms correlated with these photons always correspond to an undistorted interferometer with high-contrast fringes. But as $\lambda/d$ becomes small and the image becomes resolved, photon detection at $x_\gamma=0$ becomes extremely unlikely.

\section{Decoherence and Disturbance}
Intuitively, it seems that we should also be able to think of decoherence as resulting from some type of disturbance. But disturbance to what? Following Feynman, we might imagine that fringe washout occurs if the photon environment sufficiently disturbs the atom. Alternatively, with the commonly taught collapse postulate in mind, we might guess that the atom must disturb the environment enough to allow a which-path measurement. Understanding the centrality of entanglement in decoherence reveals that these two directions of disturbance are always mutual~\cite{stern1990phase, busch2006complementarity}. Before scattering, we could speak separately of an atom state and a photon state. But after scattering of a short-wavelength photon, we must speak of a single entangled atom-photon state. For the atom, \eq{joint} shows that it still has equal parts $\ket{\text{L}}$ and $\ket{\text{R}}$ character, but the new entangled Hilbert space vector is fundamentally different from the original non-entangled vector of \eq{psii}. For the environment, the photon has received a momentum change and also become entangled.

By evaluating $\braket{\gamma_\text{L}}{\gamma_\text{R}}$ in \eq{P}, we are considering the disturbance the atom makes to the environment. This is often the simplest way to calculate the degree of decoherence. However, we are usually interested in telling the story of the atom, so we also want to understand how it is disturbed by the photon environment. A feature of quantum mechanics now makes a dramatic appearance. Because of the entanglement, the picture of how exactly the atom is disturbed varies wildly for different choices of photon basis. 

Consider the fully decohered case $\lambda \ll d$. Using the basis of \eq{chiP}, an atom corresponding to a photon with definite $\vec{k}_\text{f}$ has undergone the disturbance
\beq
\label{kickDist}
\psi_\text{i}(\vec{r})=\ntwo\Bigl(\psi_\text{L}(\vec{r})+\psi_\text{R}(\vec{r})\Bigr)\rightarrow e^{i \kappa_x d} \psi_\text{i}(\vec{r})
\eeq
This is a \emph{momentum kick} disturbance. An atom associated with one of these photons is in a two-path state. If we detect a scattered photon with some $\vec{k}_\text{f}$, then the correlated atom we are left with was disturbed in this way. This definite momentum kick can be called ``quantum back-action," meaning it is the net effect we observe as a result of entanglement and subsequent photon detection. If there is no photon detector, then we are still free to use \eq{kickDist} to describe the disturbance to the atom. We would imagine decoherence as arising from averaging over full-contrast fringe patterns corresponding to different momentum kicks, received while the atom is traversing both paths. 

On the other hand, using the basis of \eq{joint}, the photons originate from one slit or the other. For a photon originating from the right slit, the atom disturbance looks like
\beq
\label{locDist}
\psi_\text{i}(\vec{r})=\ntwo\Bigl(\psi_\text{L}(\vec{r})+\psi_\text{R}(\vec{r})\Bigr)\rightarrow \psi_\text{R}(\vec{r})
\eeq
This is a \emph{localization} disturbance, which has no classical analogy! An atom associated with one of these photons is in a single-path state. If we detect a scattered photon with some definite origination point, then the correlated atom we are left with has been disturbed into a definite single-path state. Here, the quantum back-action is a localization. (Notice that the localization back-action we have predicted is strongly reminiscent of wavefunction collapse in a position measurement, but we have not used a collapse postulate.) If there is no photon detector, then we are still free to use \eq{locDist} to describe the disturbance to the atom. We would imagine the decohered pattern as arising from averaging over the two possibilities for single-path traversal.

Using semiclassical reasoning, Feynman's only choice was to describe fringe washout as arising from a recoil disturbance to the atom. Neglecting entanglement, we cannot not arrive at the localization-disturbance picture. This limitation becomes particularly interesting for decoherence experiments in which a semiclassical recoil does not exist~\cite{scully1991quantum, durr1998origin, busch2006complementarity}. We shall also see that the localization picture of disturbance, with its associated single-path states, is the relevant one for understanding emergence of classical behavior in complex environments.

\section{A Sketch of Classical Emergence}
Alice, Bob, and Carol are undergraduate researchers in a lab with a light microscope apparatus. Alice has not learned any quantum mechanics, and she thinks each atom will go through one slit or the other. Bob has taken an introduction to quantum mechanics, and he thinks that each atom will go through both slits. Carol has completed a full quantum course, where she learned about entangled spin pairs. She recognizes that the light microscope contains a more dramatic example of the same concept, and she knows that the character of the atom cannot be separated from the character of the photon. Carol claims that an atom passes through one slit or the other if it is correlated with a photon having a definite origination position (\eq{joint},) while the same atom would pass through both slits if it were instead correlated with a photon with definite momentum (\eq{chiP}.) 

They first perform an experiment with the light intensity set to zero, and the atom ensemble forms an interference pattern. Here, both Bob and Carol’s narratives are viable, and Alice’s classical narrative is proven wrong.

Next, they use $\lambda \ll d$ light but leave the intensity low, so that on average one photon is scattered. With no photon detection, they see a decohered atom ensemble pattern, which all three can explain. Alice explains it as a sum of single-path patterns, Bob explains it as a sum of two-path patterns with different recoil kicks, and Carol thinks that both explanations are correct. When they build atom patterns correlated with photons landing at $x_\gamma=d/2$ on the imaging detector, they see a right-shifted atom pattern, and only Alice’s and Carol’s narratives are viable. When they do similarly with the far-field detector, they construct a phase-shifted high-contrast atom fringe pattern, and only Bob’s and Carol’s narratives are viable. 

Finally, they turn up the intensity so that on average two photons are scattered from each atom. Using the single-path and definite-origin bases we have 
\beq
\label{singleBas}
\ket{\chi} \propto \Bigl( \ket{\text{L}}\ket{\gamma_\text{L}}\ket{\gamma_\text{L}} + \ket{\text{R}}\ket{\gamma_\text{R}}\ket{\gamma_\text{R}} \Bigr).
\eeq
Using the two-path and definite-momentum bases we have
\beq
\label{twoBas}
\ket{\chi} \propto \sum_{mn} a_{mn} \ket{\psi_{mn}}\ket{\gamma_{\vec{k}_m}}\ket{\gamma_{\vec{k}_n}},
\eeq
where the subscripts denote the scattering directions of the photons, and 
\beq
\label{twopath}
\ket{\psi_{mn}}\propto \Bigl(\ket{\text{L}}+e^{i \alpha_{mn}}\ket{\text{R}} \Bigr)
\eeq
is a recoiling two-path atom state. If the detectors were each 100\% efficient, then the situation would be much the same as when only one photon was scattered. In particular, Bob could use the far-field detector to determine the summed momentum kick, and once again construct a phase-shifted high-contrast atom fringe pattern. However, in practice the photon detectors do not cover the full area, so each rarely collects both photons scattered from any atom. With only one of two photons collected, we have 
\beq
\label{singleBasX}
\ket{\chi} \propto \Bigl( \ket{\text{L}}\xcancel{\ket{\gamma_\text{L}}}\ket{\gamma_\text{L}} + \ket{\text{R}}\xcancel{\ket{\gamma_\text{R}}}\ket{\gamma_\text{R}} \Bigr), 
\eeq
and
\beq
\label{twoBasX}
\ket{\chi} \propto \sum_{mn} a_{mn} \ket{\psi_{mn}}\xcancel{\ket{\gamma_{\vec{k}_m}}}\ket{\gamma_{\vec{k}_n}},
\eeq
where the crossed-off terms belong to the undetected photon. 

In the previous single-photon experiment, Alice's classical narrative failed to explain some results. However, in this experiment with imperfect detectors and a more complicated environment, her classical narrative succeeds at explaining all easily made observations. The imaging configuration allows Alice to detect only one of the two photons (\eq{singleBasX}) and then correctly predict whether the atom will land toward the left or right. She literally ``sees" atoms going through one slit or the other, and the atom detector tells a confirming story. In principle, Bob could wait to accumulate enough rare two-photon strikes on the far-field detector to once again construct a fringe pattern. But even in this relatively tame two-photon environment, proving two-path traversal is far more difficult than proving single-path traversal. If the environment were made yet slightly more complicated, for instance increasing the mean photon number~\cite{kokorowski2001single} or sending photons in from different directions, proving single-path traversal would remain easy, while proving two-path traversal would become increasingly difficult or impossible. 

\begin{figure}[htbp!]
    \centering
    \includegraphics[width=0.45\textwidth]{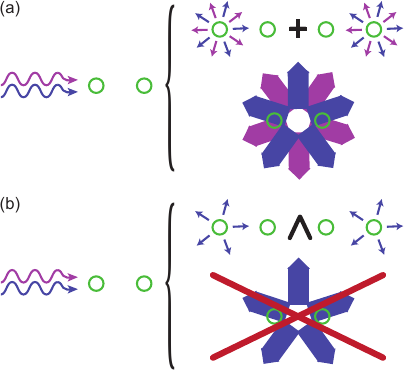}
    \caption{Two photons scatter from an atom exiting both slits. On the right, thin lines represent photons scattered from one path or the other, and thick lines represent definite-$\vec{k}_\text{f}$ photons scattered from both paths. (a) If both photons are kept in the state description, then the single-path and two-path bases are equally useful. In a laboratory, we could use either basis to make predictions, provided we detected both photons with the appropriate configuration. (b) If one photon is discarded from the state description, then the single-path basis becomes superior. In a laboratory where we only detect one photon, the single-path basis is useful for making predictions, but the two-path basis is not. With one photon lost, the atom is described as having decohered into an incoherent superposition of single-path states. (See \sec{decoherence} for elaboration.)}
    \label{cartoon}
\end{figure}

All observations in messy environments are consistent with the classical notion that atoms go through one slit or the other. 
Carol's correct entanglement narrative has no better predictive power than Alice's incorrect but simpler classical narrative. Alice's part of Carol's narrative becomes the only useful one. \fig{cartoon} shows a conceptual depiction.

\section{Pointer States}
One might ask, ``What is all the fuss? We already expected that atom interference fringes would be delicate, so that two-path traversal would be more difficult to prove." But that question cheats by embedding our classical intuition. It is far from obvious that quantum mechanics, which gives us freedom to represent states in any basis we like, should pick out \emph{any} set of favored states, much less the classical ones. Why should Alice's incorrect single-path narrative be successful in complex environments, while Bob's incorrect two-path narrative is not?

Alice, Bob, and Carol found that the basis of single-path states was the only useful one in a complex environment. A two-path description was still legal, but it was also useless for making specific predictions about atom behavior. What our three researchers have stumbled upon are called the ``pointer states" of the environment. Their name is in reference to their potential correspondence to a readout pointer on a macroscopic detector. In this case, the pointer states are the single-path states.

Decoherence theory elucidates a formal way to find pointer states for different physical conditions~\cite{zurek1981pointer, zurek1982environment, zurek1991decoherence, schlosshauer2019quantum}. The basis of pointer states is the only one useful when we are ignorant of some aspects of the environment. The wonderful thing is that these special states also turn out to be the most classical ones. It is very important to understand that we do not put in by hand our expectation for classical outcomes. The environmental pointer states, and their associated classical storylines, emerge naturally from unbiased and unmodified quantum mechanics. There are three interrelated 
ways in which pointer states distinguish themselves. 

First, pointer states are the most robust against environmental perturbation. In our simple model of \eq{oneslitchange}, the single-path states do not entangle with the photon environment, whereas two-path states do entangle. Pointer states' robustness against environmental perturbation can also be described in terms of disturbance. Single-path states experience no disturbance. In contrast, the disturbance phase-scrambles two-path states into one another if we are thinking in the definite-momentum photon basis, or it localizes two-path states into single-path states if we are thinking in the definite-origination-location photon basis.

Second, pointer states imprint their information on the environment most redundantly and robustly~\cite{zurek2009quantum,schlosshauer2019quantum}. We already encountered a closely related idea in the previous section, but we say it differently here. An atom traveling a single path imprints information about its state onto each scattered photon independently. Collecting one short-wavelength photon is sufficient to know which path the atom took. But an atom traveling two paths imprints its state information distributed amongst all scattered photons. Collecting one photon is not sufficient to know the coefficients of the recoiling two-path state, or equivalently the phase in \eq{twopath}. 

Third, pointer states do not interfere with one another after decoherence has set in. (As discussed in Sec.~\ref{decoherence}, the reduced density matrix is diagonal in the basis of pointer states.) A set of simple examples elucidate how pointer states are special in this regard. Consider the two-path states $\ket{\pm}\propto(\ket{\text{L}}\pm\ket{\text{R}})$. We saw in \eq{P} that $\ket{+}$ decoheres into non-interfering $\ket{\text{L}}$ and $\ket{\text{R}}$ states. Now consider an atom starting in state $\ket{\text{L}}$. It is perfectly legitimate to describe this state as $\ket{\text{L}}\propto(\ket{+}+\ket{-})$. In this basis, localization around the left slit occurs because $\ket{+}$ and $\ket{-}$ have opposite-phased components at the right slit which interfere to cancel each other. Now, the interaction in \eq{oneslitchange} is such that the state $\ket{\text{L}}$ persists through multiple scattering events. This means that when viewed in the $\ket{\pm}$ basis, the stable localization of a decohered $\ket{\text{L}}$ state can arise only because the $\ket{\pm}$ states keep on interfering. Generalizing, it cannot be said that any initial state decoheres into non-interfering two-path states. But it can be said that any initial state decoheres into non-interfering single-path states. Decoherence into non-interfering pointer states is called ``branching." 

Representing an atom in a two-path basis, such as $\ket{\pm}$, never becomes wrong. But it does become useless in complex environments. Without full knowledge of the environment, photon detection cannot be used to deduce that the atom traversed the slits in any of the non-classical two-path states. In contrast, the pointer-state basis is always useful. Even with environmental ignorance, photon detection can be used to deduce which of the pointer states, $\ket{\text{L}}$ or $\ket{\text{R}}$, was used by the atom to traverse the interferometer. Thus, we see the emergence of some aspects of classical physics. Our environmental ignorance limits us to making predictions and measurements for the special set of pointer states, which evolve in time following non-interfering trajectories. 

\section{Decoherence and Collapse}
\label{decoherence}
Decoherence succeeds in predicting that in complex environments the ``menu"~\cite{zurek1991decoherence} of verifiable facts is the classical one. This is cause for hope that classical physics might fully emerge from quantum mechanics, without needing to put in a microscopic/macroscopic split as did the Copenhagen theory. But there is still a gap. We have not yet seen a mechanism for an atom which started in a coherent left-right superposition to ``choose" one or the other for its decohered state, i.e. we have not yet seen a mechanism for real or apparent wavefunction collapse. 

Decoherence is not something that happens to the atom. Rather, it happens to our description of the atom when we have lost track of part of the underlying coherent entangled state. The standard tool to describe a partially or fully decohered state is the density operator. For visualization purposes, it is often expressed as the density matrix. The density operator expressing the coherent state $\ket{\chi}$ in \eq{joint} is $\rho_\chi=\ket{\chi}\bra{\chi}$. Its density matrix is 
\beq
\rho_\chi = \frac{1}{2}
\begin{pmatrix}
1 & 0 & 0 & 1\\
0 & 0 & 0 & 0\\
0 & 0 & 0 & 0\\
1 & 0 & 0 & 1
\end{pmatrix}
\eeq
in the $\ket{\text{L}}\ket{\gamma_\text{L}}$, $\ket{\text{L}}\ket{\gamma_\text{R}}$, $\ket{\text{R}}\ket{\gamma_\text{L}}$, $\ket{\text{R}}\ket{\gamma_\text{R}}$ basis. The diagonal terms are the ``populations," and the off-diagonal terms are the ``coherences." For our discussion, it is not necessary to understand anything about the density matrix other than two things. First, the populations always add to 1, meaning that we have not lost atoms or photons. Second, the non-zero off-diagonal terms mean that the atoms states of this basis ($\ket{\text{L}}$ and $\ket{\text{R}}$) are capable of exhibiting interference effects under the right circumstances. And \fig{partials} confirms this; if we have access to a far-field photon detector, we can construct full-contrast atom interference fringes, even for $\lambda \ll d$. 

Now, to describe the case that we do not collect photons, we can use $\rho_\chi$ to calculate (by tracing out the photon states, i.e. summing over them) the corresponding ``reduced" density matrix for the atom. For $\lambda \gg d$, we obtain
\beq
\label{rhoAlong}
\rho_\text{A} = \frac{1}{2}
\begin{pmatrix}
1 & 1 \\
1 & 1
\end{pmatrix}
\eeq
in the $\ket{\text{L}}$, $\ket{\text{R}}$ basis. For $\lambda \ll d$, we obtain
\beq
\label{rhoAshort}
\rho_\text{A} = \frac{1}{2}
\begin{pmatrix}
1 & 0 \\
0 & 1
\end{pmatrix}.
\eeq
Again the populations add to 1, but whether or not there are coherences depends on whether the photon can resolve the slit separation. The vanishing of off-diagonal terms in \eq{rhoAshort} says that, without a photon detector, we cannot observe interference between $\ket{\text{L}}$ and $\ket{\text{R}}$ states if $\lambda \ll d$. Using the density matrix adds nothing new to what we have already covered in \eq{P} and the subsequent discussion, but it does present the results concisely.

Now, here is a point of confusion which is important to clarify. Density matrices can describe ensembles as well as individual particles, and a diagonal density matrix like \eq{rhoAshort} is often called a ``mixed state." That term makes perfect sense if we are using it to describe a statistically mixed ensemble of definite-state atoms. But here we are talking about a single atom. Does a diagonal density matrix mean that the atom wavefunction collapsed into either $\ket{\text{L}}$ or $\ket{\text{R}}$, with us being ignorant about which one? No, at least not in the ``realist" view that the wavefunction describes the microscopic reality of the atom. The original entangled state $\ket{\chi}$ had equal parts $\ket{\text{L}}$ and $\ket{\text{R}}$ character, which we can verify by using detected photons to reconstruct atom fringes (\fig{partials}.) The atom does not lose half its character just because we ignore the photon.

For wavefunction-realists, decoherence most certainly does not mean collapse. Yet we cannot observe interference. It could be said that our environmental ignorance forces us to describe the atom as being in an ``incoherent superposition." That is non-standard language, but it is instructive. (The more standard way of saying exactly the same thing is to insist that each atom by itself is described by a decohered density matrix like \eq{rhoAshort}, regardless of whether it is part of an ensemble.) We can write the incoherent superposition as 
\beq
\label{psiDec}
\boldsymbol{|}\psi\} = \frac{1}{2} \ket{\text{L}} \land \frac{1}{2} \ket{\text{R}}.
\eeq
That is also non-standard notation, and $\boldsymbol{|}\psi\}$ does not refer to a vector in Hilbert space. But this shorthand is helpful because it reminds us that the atom state has decohered but not collapsed into one state or the other. \fig{cartoon} shows a graphical depiction of decoherence arising from two photons being scattered but only one detected. 

That all sounds harmless enough when we are talking about microscopic atoms and photons. However, unitary evolution of \ss\ continues to hold true when tested on much larger systems~\cite{pedalino2025probing, kotler2021direct, hornberger2003collisional, hackermuller2004decoherence}. (Indeed, large-scale quantum computation would require this winning streak to continue.) There seems to be every reason to expect unitary evolution will continue to hold true up to macroscopic scales. But then the implications for wavefunction-realists become strange.

Consider, the light microscope, with multiple photons scattered and only a fraction collected in the imaging detector. Since we have lost photons, we must describe the atom as having decohered (\eq{psiDec}.) Say that our photon detector is engineered such that a mechanical readout arrow swings either left or right when a photon is detected in the cluster of pixels near $x_\gamma=-d/2$ or $x_\gamma=+d/2$. The detector apparatus is just a collection of atoms following physical laws, so we can describe the arrow states by \ket{\,\includegraphics[scale=.4,valign=c]{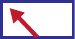}\,} and \ket{\,\includegraphics[scale=.4,valign=c]{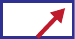}\,}. (Each of those kets is understood to represent any one of a large number of microscopic configurations which are macroscopically indistinguishable.) Photon detection results in atom-detector entanglement, and if unitary evolution continues to work up to these scales, the atom-detector state evolves into
\beq
\label{mwi}
\boldsymbol{|}\chi_{\text{AD}}\} = \frac{1}{2} \ket{\text{L}}\ket{\,\includegraphics[scale=.4,valign=c]{detL}\,} \land  \frac{1}{2}\ket{\text{R}} \ket{\,\includegraphics[scale=.4,valign=c]{detR}\,}.
\eeq
We can also say the macroscopic detector is in the incoherent superposition $\boldsymbol{|}\psi_\text{D}\} = \frac{1}{2} (\ket{\,\includegraphics[scale=.4,valign=c]{detL}\,} \land \ket{\,\includegraphics[scale=.4,valign=c]{detR}\,})$. But it is \emph{not} in the coherent superposition $\ket{\psi_\text{D}} = \ntwo (\ket{\,\includegraphics[scale=.4,valign=c]{detL}\,} + \ket{\,\includegraphics[scale=.4,valign=c]{detR}\,})$, because that erroneous expression neglects entanglement with the atom and many other degrees of freedom. 

A biological organism, such as a cat, looking at the detector would be carried along into an incoherent superposition of having two different neurological experiences. The decohered detector-cat system would be $\boldsymbol{|}\chi_\text{DC}\} = \frac{1}{2} (\ket{\,\includegraphics[scale=.4,valign=c]{detL}\,}\ket{\,\includegraphics[scale=.1,valign=c]{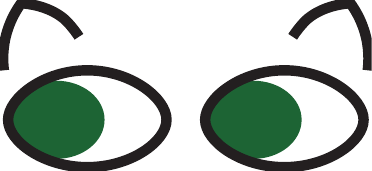}\,} \land \ket{\,\includegraphics[scale=.4,valign=c]{detR}\,}\ket{\,\includegraphics[scale=.1,valign=c]{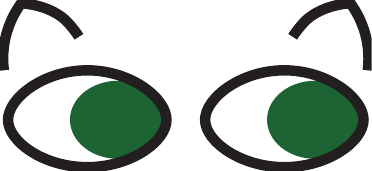}\,} )$. Reality now contains two different versions of the cat, each experiencing a definite measurement outcome. There has been no actual collapse, but each version of the cat thinks there has been. As we have seen, plain quantum mechanics predicts that a physicist experimenting on the the atom-detector-cat system will be incapable of verifying the actual existence of the incoherent superposition. We cannot even see interference fringes of the decohered atoms, much less of the decohered cats.

If all matter is governed by plain quantum mechanics, then wavefunction-realists are forced to accept that macroscopic reality is constantly and unavoidably splitting in this way, because of the superpositions which propagate up from microscopic reality. This is the Many-Worlds interpretation (MWI) of quantum mechanics, originally described in the 1957 doctoral thesis of Hugh Everett~\cite{everett1957hugh, everett1957relative}. Each evolving classical storyline of the incoherent superposition is often called a ``branch," but the term ``world" can also be used. Contrary to a popular misconception, these ``worlds" are not put in by hand; they emerge unavoidably from persistent unitary evolution, decoherence, and a realist's insistence that the wavefunction describes a reality outside of our minds. (See~\cite{tegmark2007many, carroll2020something} for accessible discussion of other features of MWI, such as apparent randomness, whether all branches are equally real, and natural extension to quantum field theory.)

Not everyone thinks that the realist interpretation is best. ``Epistemic" interpretations do not dispute that MWI makes correct predictions. Rather, they are motivated by a sense that MWI is too weird to be believable, or that it suffers from other less obvious philosophical problems. Epistemic approaches claim that instead of describing any microscopic reality, wavefunctions are only mathematical tools for predicting experiment outcomes. These interpretations represent a departure from how most of us think science. But perhaps most of us are wrong in general, or perhaps quantum mechanics presents a special case where non-realist thinking is required. In the QBism epistemic interpretation~\cite{fuchs2000quantum, mermin2012commentary}, each observer has their own personal version of the wavefunction describing some physical system, and this wavefunction is updated once new information becomes available. Once the observer learns which of the decohered branches predicted by unitary evolution corresponds to their experience, their version of the wavefunction collapses.

MWI and epistemic approaches are pure interpretations of plain quantum mechanics; with the help of decoherence theory, they remove the nonsensical microscopic/macroscopic split of the Copenhagen interpretation, without adding anything extra or modifying equations. But some physicists and philosophers find neither interpretation credible, and there are plenty of proposed modifications to quantum theory. Some realist theories include an objective collapse by adding non-linear stochastic terms to \ss~\cite{bassi2003dynamical}. Here, unitary evolution is correct on short timescales, until the new dynamics suddenly collapses the massively entangled wavefunction. The de Broglie-Bohm pilot wave theory~\cite{bohm1952suggested} is another realist modification which removes collapse entirely by adding  to the theory non-local hidden variables. These are definite positions for the particles, which move deterministically in a potential given by the differently-interpreted wavefunction.

\section{Conclusion}
Decoherence is the penalty we pay for trying to separate out one part of a non-separable entangled state. Wonderfully, this ``illegal" separation accomplishes what Copenhagen clumsily attempted by imposing its shifty split. The quantum freedom to choose different bases creates an ambiguity as to the facts of any unfolding story. In Feynman's light microscope, is the atom following a classical single-path trajectory or a non-classical two-path trajectory? With full environmental information, both narratives are legitimate and both make experimentally verifiable predictions. But without ever imposing a microscopic/macroscopic distinction, decoherence causes plain quantum mechanics to pick out the classical single-path storylines as the only ones which can be recorded and used to make predictions in a complex environment. These decohered storylines each contain a set of unambiguous and causally connected facts. 

The other thing that the Copenhagen split accomplished was to predict a single outcome of any macroscopic measurement. We now understand the branching of decoherence accomplishes this naturally, without imposing a split. Unitary evolution predicts a single detector outcome on each decohered branch, with different outcomes on different branches. But at this point it is possible to say very different things about what this means for macroscopic reality, depending on how we interpret the microscopic wavefunction.

The traditional realist interpretation of science is that solutions to physical equations describe physical reality. For example, the trajectory of a projectile launched into outer space continues to describe something real, even if it is never again observed. When applied to the wavefunction of unmodified quantum mechanics, the realist conclusion is Everettian MWI. Microscopic superpositions unavoidably percolate up to macroscopic superpositions, and each moment in time is branching forward into parallel decohered macroscopic realities. Some branches include our existence while others do not, but all predicted branches are real. 
Decoherence also plays a second role here. We are unable to observe interference between decohered branches, so we are incapable of proving that branches other than ours actually exist. The wavefunction never actually collapses, but decoherence causes the illusion of collapse.

Alternatively, epistemic interpretations understand the wavefunction to be only a tool for making predictions. If the wavefunction never described microscopic reality, there is no basis for saying that any of the decohered branches predicted by unitary evolution exist, other than the one we are experiencing. 

As a practical matter, MWI and epistemic approaches use identical equations, naturally include decoherence in identical ways, and make identical predictions for experiments. Opinions differ on whether the distinction is therefore an unimportant matter of philosophical preference~\cite{tegmark2007many, camilleri2009history}. However, MWI and epistemic theories are together testable. For instance, if interference fringes of a big object were found to wash out for an inexplicable reason, then their underpinning unitary evolution would be falsified. 

It is of course possible that quantum mechanics, and its extension quantum field theory, need to be modified in some fundamental way. Advocates point out that quantum theory has not yet been reconciled with gravity, and that there are still orders of magnitude between where it has been tested and the scale of macroscopic detectors. Alternate theories deserve to be tested when possible, but we should keep in mind that unitary evolution of \ss\ has continued to prove successful under increasingly stringent conditions~\cite{pedalino2025probing, kotler2021direct, hornberger2003collisional, hackermuller2004decoherence}. Quantum mechanics has no crisis which proposed modifications would address, other than our human crisis of being forced to question cherished notions of reality or science.

Half a century ago, physicists could believe that wavefunctions described the reality of small things and that there was only one macroscopic reality. But after the development of decoherence theory and experimental scrutiny at relatively large mass scales, we can no longer rely on wavefunction collapse to provide that comfortable scenario. We must either say that plain quantum mechanics correctly predicts detector clicks without describing microscopic reality, or that it predicts a vast number of macroscopic realities besides the one we are currently experiencing. 

\begin{acknowledgments}
We gratefully acknowledge stimulating and clarifying discussions with Isaac Cross, Jerry Gabrielse, Tim Kovachy, Jay Lawrence, and Bill Wootters.
\end{acknowledgments}
        
\appendix
\section{Calculation of Interference Patterns}
\label{appCalc}
We are interested in obtaining atom patterns for a detector closer than the far-field limit, such that the single-slit patterns are somewhat separated from one another. To accommodate this requirement, we use a path integral approach. The phase acquired by an atom momentum eigenstate traveling from a source point $(x', y', 0)$ to a detector point $(x,0, L)$ is given by 
\beq
\Phi(x,x',y')=\frac{p_x}{\hbar} (x-x') - \frac{p_y}{\hbar} y' + \frac{p_z}{\hbar} L - \omega t,
\eeq
where $\omega=(p_x^2 + p_y^2 + p_z^2)/(2 m \hbar)$, and $t$ is the propagation time. (Note that the presence of recoil has turned the problem, for the moment, from a 2D one to a 3D one, and a range of $y'$ sources can contribute to detection at $y=0$.) If we use the classical expressions $p_x/p_z = (x-x')/L$, $p_y/p_z = -y'/L$, and $t=m L/p_z$, then after subtracting off common phase contributions, we arrive at  
\begin{align}
\Phi(x,x',y') &= \frac{p_z} {2 \hbar L} \left((x-x')^2+(y')^2\right)\\
&= \frac{p_0+\hbar k_z} {2 \hbar L} \left((x-x')^2+(y')^2\right). \nonumber
\end{align}
The $k_z$ term is of order $\hbar^2 k L/(p^2_0 a^2)$ and the $y'$ term is of order $\hbar k^2 L/p_0$. By increasing $p_0/(\hbar k)$, these terms can be made arbitrarily small. In other words, for sufficiently short transit times, the photon scattering perturbation has negligible effect on the phase accumulated over transit. We use this high-momentum limit to simplify the calculations, and we have
\beq
\label{Phi}
\Phi(x,x') = \frac{p_0} {2 \hbar L} (x-x')^2.
\eeq

The wavefunction $\psi(x)$ at some point on the detector is found by integrating the contributions from all source points $x'$. For this purpose, the use of classical trajectories, used to derive \eq{Phi}, can be easily justified for $p_0 \gg p_x$~\cite{sanz2015full} or more generally by using path-integral theory~\cite{storey1994feynman}. If there is no photon scattered, the wavefunction shares a common phase at all points on the apertures, and we have
\beq
\label{psi0}
\psi(x) \propto \int_\text{M} \text{d}x' e^{i\Phi(x,x')},
\eeq
where the subscript M denotes integration over the aperture mask. If there is a photon scattered, then for the atom state correlated with photon state $\ket{\gamma_n}$, \eq{psi0} becomes
\beq
\label{psigamma}
\psi_n(x) \propto \int_\text{M} \text{d}x' \braket{\gamma_n}{\gamma_{x'}} e^{i\Phi(x,x')},  
\eeq
where $\ket{\gamma_{x'}}$ is a photon emitted from $x'$. 

Expressions for the matrix elements to be used in \eq{psigamma} are derived in Ref.~\cite{tan1993loss}. For atom wavefunctions correlated with photon imaging-configuration detection at $x_\gamma$, we use their Eq.~49 to find $\braket{\gamma_{x_\gamma}}{\gamma_{x'}}$. For atom wavefunctions correlated with scattered photon direction, we have
\beq
\label{kfProj}
\braket{\gamma_{\vec{k}_\text{f}}}{\gamma_{x'}}=S(\vec{k}_\text{f}) e^{i \kappa_x x'},
\eeq
where $S$ is the scattering amplitude for a photon scattered from $x=0$. 

Calculation of the atom patterns corresponding to no photon detection can be done by averaging over either photon basis. We use the $\ket{\gamma_{\vec{k}_\text{f}}}$ basis. After integrating over the azimuthal angle about $\hat{x}$, the joint probability density becomes~\cite{tan1993loss}
\beq
\label{P0}
\mathcal{P}_{x, \kappa_x} \propto (2k^2 + \kappa_x^2-2 k \kappa_x) \left|\int_\text{M} \text{d}x' e^{i\kappa_x x'}e^{i\Phi(x,x')}\right|^2.
\eeq
This equation corresponds to Eq.~53 of~\cite{tan1993loss}, whose procedure we followed, but we have generalized their expression to accommodate our finite aperture width and non-far-field atom detection. We numerically integrating over $k_{\text{f}x}$ to find the atom distribution $\mathcal{P}_x$. Since we only consider $\lambda \gg a$, we simplify the computation by treating the recoil term as constant over each slit.

Finally, narrow-slit expressions for interferometer phase, used in App.~\ref{appSC}, follow from the above expressions. For a photon with definite $\vec{k}_\text{f}$, we use \eq{kfProj} in Eq.~\ref{psigamma}. The phase difference between narrow slits is $\Delta \phi = \phi_\text{L} - \phi_\text{R}$, with the sign chosen to make the interferometer phase increase as $x$ increases. Using $\phi_\text{L,R} = (p_0)/(2 \hbar L) (x \pm d/2)^2 \mp \kappa_x d/2$, we get
\begin{align}
\label{phiz}
\Delta\phi&=\frac{p_0 d}{L\hbar} x -\kappa_x d\\
\label{phix}
&=\frac{p_x d}{\hbar}-\kappa_x d. 
\end{align}
If no photon is scattered, we have
\begin{align}
\label{phi0z}
\Delta\phi_0&=\frac{p_0 d}{L\hbar} x\\
\label{phi0x}
&=\frac{p_x d}{\hbar}. 
\end{align}
These are the well-known expressions for the phase of an unperturbed two-narrow-slit interferometer.

\section{Semiclassical Pitfalls and Proper Calculation of Deflection Phase Change}
\label{appSC}
First, we outline a sometimes-used sloppy argument, which arrives at the wrong result despite initial appearances of success. Since recoil takes $p_x \rightarrow p_x + \hbar \kappa_x$, we might be tempted to modify the well-known Eq.~\ref{phi0x} by writing $\Delta\phi_0 \rightarrow \Delta\phi_0 + \kappa_x d$. From this expression, we could mistakenly conclude that we have arrived at the anticipated result, that randomizing recoil angles washes out atom fringes for $\lambda < d$. But it is the final momentum rather than the initial momentum which determines where the atom lands on the detector, and the above argument forgets that as we change $p_x$ we also change the detection position $x$. What we actually calculated was simply how the original unperturbed fringe phase changes with $x$ and its corresponding $p_x$. There is no predicted fringe washout. 

Continuing to make the mistake of neglecting recoil phase, we can be less sloppy than above by using Eq.~\ref{phi0z}. We might wonder whether the longitudinal component of recoil achieves the expected result. The substitution $p_0 \rightarrow p_0 + \hbar \kappa_z$ predicts a phase smearing $\delta \phi = (x/L) (\kappa_z d)$. This can be understood as transit-time broadening destroying the one-to-one correspondence between $p_x$ and $x$. This type of longitudinal decoherence is real, but it is not important for $x/L < \lambda/d$. If it were the whole story, there would still be of order $p_0/(\hbar k) \gg 1$ high-contrast fringes at small atom angles, even for $\lambda < d$. 

Finally, we properly include recoil phase to find how the post-recoil and pre-recoil interferometer phases are related. For an atom with pre-recoil transverse momentum $p_{x_0}$, the pre-recoil interferometer phase (\eq{phi0x}) is $\Delta\phi_0=p_{x_0} d/\hbar$. As discussed in Sec.~\ref{anticipating}, recoil introduces a phase difference between atom wavefunctions leaving the two slits, and the post-recoil interferometer phase is given by \eq{phix}. Recoil deflects the atom to a new angle, and using $p_x = p_{x_0} + \hbar \kappa_x$, we find $\Delta\phi = \Delta\phi_0$ because the $\kappa_x$ terms cancel. Thus for a two-slit interferometer, a deflected atom keeps its pre-recoil interferometer phase, as Feynman implied. This result predicts fringe washout for $\lambda<d$, whereas the misguided approaches neglecting recoil phase do not.

\bibliographystyle{aipnum4-2}
\bibliography{bib.bib}
\end{document}